\documentclass[prl,letterpaper, preprint]{revtex4}

\include{epsf}

\tighten

\begin{document}

\title{Neutrino-induced Collapse of Bare Strange Stars\\
Via TeV-scale Black Hole Seeding}

\author{
Peter~Gorham, John Learned, and Nikolai Lehtinen}
\affiliation{Department of Physics and Astronomy, University of Hawaii
at Manoa, 2505 Correa Rd., Honolulu, HI, 96822}


\begin{abstract}
There is increasing observational evidence for the existence of 
strange stars: ultra-compact objects whose interior consists 
entirely of deconfined quark matter. If confirmed, their existence places
constraints on the rate of formation of microscopic black holes in models
which invoke a TeV-scale Planck mass. In such models, black holes can
form with $\sim$ TeV masses through nuclear interactions of particles with 
PeV and greater energies. Once formed, these black
hole states are unstable to Hawking radiation, and rapidly decay. 
However, if such a black hole forms in the
interior of a strange star, the density is high enough that the 
decay may be counterbalanced by accretion, and the black hole can grow,
leading to subsequent catastropic collapse of the star. 
A guaranteed source of ultra-high energy
particles is provided by the cosmogenic Greisen neutrinos, as well as
by ultra-high energy cosmic rays, and the implied lifetimes for strange
stars are extremely short, contrary to observations. The
observed lifetimes of strange star candidates thus effectively
exclude Planck mass scales of less than $\sim 2$~TeV with comparable
black hole masses, for up to 2 extra dimensions. Seeding of strange
star collapse in scenarios with a larger number of extra-dimensions
or with higher mass black holes remains a possibility, and may provide
another channel for the origin of gamma-ray bursts.
\end{abstract}
\maketitle

To address the large disparity between the four-dimensional Planck 
mass ($M_4 \sim 10^{19}$~GeV) for gravitational symmetry-breaking and the 
corresponding electro-weak scale ($\sim 100$~GeV), it has been proposed
that space may contain compact extra dimensions with scales up to 
$\sim 0.1$~mm~\cite{Ark98, Ant98}. In such models, the fundamental $n$-dimensional
Planck scale $M_D$ (with $n= D-4$ extra dimensions) 
could appear at energies of O(TeV). At this fundamental
scale, gravitational interactions between particles can become
strong enough to produce a
black hole (BH) with a $\sim$~TeV mass~\cite{Arg98,Emp00}.
At these masses, the black holes decay very rapidly due to 
Hawking radiation, unless there are sufficient further nuclear interactions
within their lifetimes to offset their evaporation, requiring matter at
near nuclear densities. Such material occurs naturally only in the cores of
ultra-compact stars.

In the case of a neutron star, the density rises from $\sim 1$ gm cm$^{-2}$
to near-nuclear density~\cite{Lor93,Akm98,Lat01} 
over the first 50-200~m. 
Any incoming particle with sufficient energy to 
produce a black hole will interact well before the high density region.
The black hole
thus formed initially encounters matter at only the crustal density, of order
$10^{-6}$~to~$10^{-4}$~fm$^{-3}$ (1 fm$^{-3} = 1.67 \times 10^{15}$ gm cm$^{-3}$).
This density is inadequate to stabilize the black hole against
decay. However, if stars exist with interior densities that approach
nuclear density very close to the surface, evaporation of the black
hole may be offset by accretion of the surrounding matter. With sufficient
density, runaway growth of the black hole will lead to catastrophic 
collapse of the star.

Theoretical predictions for the existence of strange stars arose from the
hypothesis that $u,d,s$ quark matter may be the final ground state of
the strong interaction~\cite{SStar0}, 
and thus the final stable state prior to gravitational collapse. 
Recently, a number of studies have led to
several proposed candidates for such stars, among them the X-ray pulsar
Her X-1~\cite{Dey98}, 
the bulge sources 4U 1728-34~\cite{XDL99b} 
and 4U 1820-30~\cite{Bom97}, the recently discovered
ms pulsar SAX J1808.4-3658~\cite{XDL99a}, 
the X-ray pulsar GRO J1744-28~\cite{Che98},
and the isolated neutron star RX J185635-375~\cite{Dra02, Pon02}. 
Although further refinements
of neutron star models may eventually fit some of these candidates,
the observations favor masses and radii which are presently inconsistent
with any neutron star model.

A distinguishing feature of such stars is that their mass-radius
relation behaves in a manner opposite to that of classical neutron stars:
whereas a neutron star's radius decreases with increasing mass, a strange
star, by virtue of its uniform maximal nuclear density, 
must increase its radius
with increasing mass~\cite{Bom02}. 
It has been proposed that neutron stars may in
fact be ``seeded'' with quark matter to convert 
them to strange stars~\cite{Alc86}.
In this letter, we consider a more radical seeding
process: that of conversion of a quark star to a black hole by a
TeV-mass black hole seed.

Formation of TeV-scale black holes is qualitatively a result of 
high energy collisions in which two particles pass within a
Schwarzschild radius $r_s$ of each other with a center of mass energy
of $E_{cm} = \sqrt{s}$ where for a nucleon at rest $s = 2m_N E_i$ 
with $m_N \simeq 1$~GeV and
$E_i$ the lab-frame energy of the incident particle. The cross section
for this process 
is predicted to be approximately geometric with 
$\sigma_{\rm BH} \simeq \pi r_s^2$. The implied energy threshold for
production on nucleon targets at rest is 
$E_{i,thr}~=~5~\times~10^{14}~{\rm eV}~(M_{\rm BH}~/~{\rm 1~TeV})^2$~.
It is evident that this threshold  
is at present out of reach of any fixed target accelerator.
However, the Large Hadron Collider (LHC) has a goal of $\sqrt{s} = 14$~TeV,
corresponding to $E \simeq 10^{17}$~eV for interactions with
a fixed target. Thus production of black holes for
$M_{\rm BH} \sim M_D$ up to several TeV may be within its reach~\cite{Emp00, Dim01, Gid02}. 
This possibility has generated considerable recent interest.

There are several cosmic sources of particles of energies above
0.1 EeV which could play a role in black hole production.
Two sources of particular interest 
are the high energy cosmic rays, and the 
related cosmogenic neutrino flux~\cite{Eng01}, which arises
as a result of the integrated interactions of the highest energy
cosmic rays (with energies above 
$\sim 3 \times 10^{19}$~eV) with the cosmic microwave
background radiation throughout the universe, the so-called
Greisen-Zatsepin-Kuzmin (GZK) process~\cite{GZK}. 
The GZK neutrinos have been identified already as a
likely source of black-hole production in existing neutrino detectors,
and useful limits on the black-hole production cross section have
already been derived from air shower data~\cite{Anch02b, Tu02}. 

We  stress here in passing that
the flux of GZK neutrinos is a generic result of the GZK
cutoff, and is predicted in varying quantities for all standard
models of ultra-high energy cosmic-ray propagation, as well as for many models
in which the GZK cutoff is violated.
The only assumptions necessary for the the existence of
a GZK neutrino flux are that
(a) the local universe is not greatly different from any other 
cosmic locale in its energy density of $\geq 3 \times 10^{19}$ ~eV cosmic
rays; and (b) that photopion production is well-behaved at
the center-of-momentum-frame energy ($\sim 1$~ GeV) 
needed for the GZK process. Since cross sections are well-known at
GeV energies, this latter condition reduces to a requirement on 
the accuracy of the Lorentz transformation for $\gamma \simeq 10^{11}$.

The crucial factor in determining the likelihood of
black hole production by a given energetic particle is the black-hole
production cross-section. Estimates are based on geometric
arguments for interactions partons are based
on the $D+n$-dimensional Schwarzshild radius~\cite{Dim01}:
\begin{equation}
\sigma_{\rm BH} \simeq \pi r_s^2 = {1 \over  M_D^2 }
\left [ {M_{\rm BH} \over M_D}  
\left ( { 8~ \Gamma({n+3 \over 2}) \over n+2 } \right ) 
\right ]^{2\over n+1}~.
\end{equation}
Extensions of this argument for 
high energy neutrino interactions~\cite{Feng02}
gives cross sections of order $10^{-32}$ to $10^{-30}$ cm$^{2}$
for neutrinos from 0.1-1 EeV~\cite{Anch02a, Anch02b, Alv02, Feng02}. 
Evaluation of production cross sections
for $\sim$~few TeV black holes in $p \bar{p}$-interactions
at the LHC indicate cross sections in the range
of $10^{-34}$ to $10^{-32}$~cm$^2$ for $n$ 
in the range of 4 to 6~\cite{Dim01}.
Black hole production in hadron collisions may thus 
be comparable to top quark production at the LHC.

There is still considerable debate over whether there is
an exponential suppression of these values, leading to cross sections
1-3 orders of magnitude smaller~\cite{Vol01}. 
In either case, it is evident that
the fraction of cosmic ray interactions which can produce black holes
will be at most the ratio of the black-hole cross section to the 
total hadronic interaction cross section, 
thus $\simeq 10^{-10}$ to $10^{-7}$ per interaction. 

For ultra-high energy neutrinos, however, the black hole production
cross section can exceed the standard model deep-inelastic hadronic
cross section by 1-2 orders of magnitude~\cite{Anch02a,Anch02b,Alv02,Feng02}, 
making black-hole production
the dominant process for neutrinos in many cases. Even for the
exponentially-suppressed cases, the fraction of EeV interactions that
can produce black holes is no less than about 10\%. Thus, although
the flux of GZK neutrinos is, for most models, less 
than that of the cosmic rays
at these energies, the high probability for black-hole production
in this scenario means that the GZK neutrino flux
will dominate the rate.

If bare strange stars exist, their density profile is expected to
be a nearly constant $10^{15}$ gm cm$^{-3}$ 
($\sim$1 fm$^{-3}$) up to within 1 fm of the surface~\cite{Glen00}. 
Under these conditions, the neutrino interaction length is
$L_\nu = (\rho_{s}~ \sigma_{\rm tot}~ N_A)^{-1} \simeq 
 10^{-8}~{\rm cm}~(10^4~{\rm pb}/\sigma_{tot})
(1~{\rm fm^{-3}}/ \rho_{s})$ where 
$\sigma_{tot}=\sigma_{\rm BH}+\sigma_{sm}$ is the sum of the black hole]
and standard model neutrino cross sections.
The interaction thus evidently takes place within a few atomic diameters
of the surface.  

The rest frame lifetime for the black hole
due to evaporation via Hawking radiation is estimated to be~\cite{Emp00} 
\begin{equation}
\label{tau}
\tau \sim {1 \over M_D} 
\left ( {M_{\rm BH} \over M_D} \right )^{(3+n) \over (1+n)}~.
\end{equation}
For $M_D \simeq M_{\rm BH} \simeq 1$~TeV, the lifetime is
$\sim 10^{-27}$~s. Evaporation in the black hole rest frame is thus extremely
rapid if there is nothing to prevent it.


The Lorentz factor of the black hole $\gamma = E_{\nu}/M_{\rm BH}$ 
has a significant effect on the black hole
evaporation process~\cite{Note1}. The black hole is created within the 
highly degenerate Fermi gas of the interior of the strange star. 
For young strange stars this consists of a 
photon component with effective temperature
of order 10-25~MeV, and a quark component with Fermi temperature
$T_F \simeq 0.4-1$~GeV, with an upper limit set by the $c$-quark 
mass at 1.15-1.35~GeV. 

The Lorentz boost modifies the angular distribution of
the thermal bath temperature by the factor 
$T'(\theta) = T_{bath} [\gamma(1 + \beta \cos \theta)]^{-1}$. Using the
Stefan-Boltzmann law (which is appropriate here since the
boosted effective temperature is much higher than the fermion
masses), this factor can be integrated over all
directions to yield the effective temperature of
the bath as observed in the black hole frame~\cite{Light75}:
\begin{equation}
T_{eff} = T_{bath} \sqrt{\gamma}~(1 + {\beta^2 \over 3})^{1 \over 4}~.
\end{equation}
The bath temperature is determined by the temperatures of each of
the partial pressure components in the thermal bath, but is
dominated by the Fermi pressure of the quark-gluon plasma.

The Hawking temperature of the black hole is given by
\begin{equation}
T_H = M_D \left ( {M_D \over M_{\rm BH} } ~
{n+2 \over 8~ \Gamma({n+3 \over 2}) } \right )^{1 \over n+1 } ~
{n+1 \over 4\sqrt{\pi}}~,
\end{equation}
and falls typically in the range of $T_H = 100-500$~GeV for 
$M_D \simeq 1-3$~TeV and similar black hole masses.
To prevent immediate evaporation of the black hole, we
require $T_{eff} \geq T_H$ which gives a condition on the 
initial neutrino energy:
$E_{\nu,thr} ~\simeq~ M_{\rm BH} ( 3T_H / 4T_{bath})^2 $.
For  $T_{bath} =  T_F$, $M_{D}\sim M_{\rm BH} \leq 5$~TeV, $n=1,2,3,4$, 
this requirement yields 
$E_{\nu,~thr} = 
(1.5 \times 10^{19},5 \times 10^{19}, 10^{20}, 1.5 \times 10^{20})$~ev.
Black holes created by neutrinos of energy greater than this will be
initially stabilized by the apparent temperature of the bath in their own
rest frame. 

The mass evolution of the initially relativistic 
black hole through accretion
can in principle also be treated by 
a quasi-thermodynamic approach. Kinematically, however, the first interaction
of the boosted black hole with a quark is the most significant, since
the black hole absorbs all of the mass energy of this particle, including
the large portion of its own boost that the particle has in the center of
momentum frame. Thus we treat the fermion accretion problem as
collisions with a black hole with a capture cross section given by the
classical value $\sigma_c = (27/4)\pi r_s^2$
which accounts for the impact parameter of 
geodesics into the hole~\cite{Kip86,Anch02b}.

\begin{figure}
\begin{center}
\leavevmode
\epsfxsize=3.55in
\epsfbox{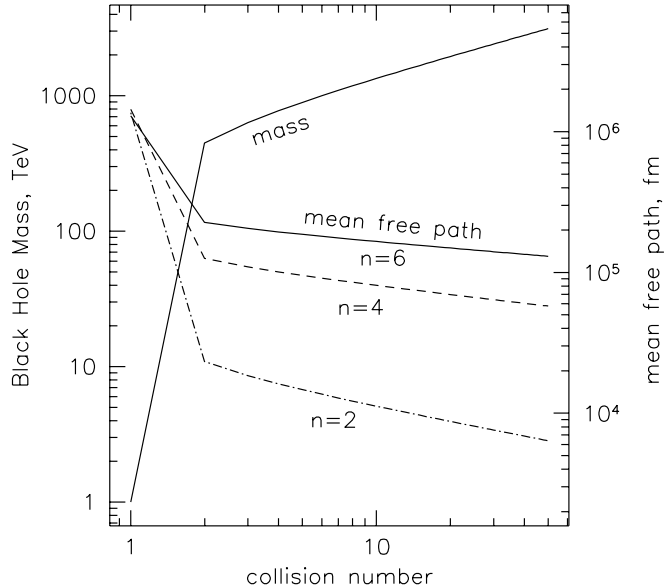}
\caption{Black hole mass \& mean free path evolution vs. number of
collisions in the interior of a strange star for $M_{\rm BH, initial}= 1$~TeV,
$E_{\nu} = 10^{20}$~eV (the black hole creation is the first 
interaction). The largest fractional mass increase
and drop in the mean free path happens at the first subsequent collision.}
\label{coll1}
\end{center}
\end{figure}

We have treated the black hole- parton accretion process both numerically
and semi-analytically, and we find that, as long as the black hole effective
lifetime is long enough for it to interact with a single parton with 
reasonable probability, the large amount of mass-energy absorbed in this first
interaction leads invariably to the black hole eventually coming to rest
with  a mass equal to the energy of the particle which created it:
$M_{\rm BH,~final} \simeq E_{\nu}$. For $E_\nu = 10^{20}$~eV,
$n=2$, and initial $M_{\rm BH} \sim M_D \sim 1$~TeV, 
this requires of order $10^{11}$ collisions with massive quarks in the
star, over a total distance of order 0.1 mm. 
A plot of the first 50 collisions of
the black hole mass evolution for this case is shown in Fig. 1.

At this stage the black hole is approximately at rest 
with a much reduced surface
temperature due to its increased mass.
If $T_F \geq T_H$ at this stage the black hole can continue to grow; if
not, evaporation will proceed again. 
This condition can be solved for the minimum
incoming neutrino energy required for runaway black hole growth:
\begin{equation}
\label{Enumin}
E_{\nu,~min}~=~ { M_D^{n+2} \over T_{bath}^{n+1} }~
{n+2 \over 8~ \Gamma({n+3 \over 2})}~
\left ( {n+1 \over 4\sqrt{\pi}}  \right )^{1/(n+1)}~.
\end{equation}
For $n=1,2,3,4$, and $M_{\rm BH} \sim M_D \sim 1$~TeV,
and $T_{bath} = 1$~GeV, $E_{\nu,~min} = (2.5 \times 10^{16},
2.8 \times 10^{19}, 3 \times 10^{22}, 4 \times 10^{25})$~eV, which
shows that there is a region of the $n \leq 2$ extra-dimension parameter
space that is probed by the expected GZK neutrino flux. 



\begin{figure}
\begin{center}
\leavevmode
\epsfxsize=3.5in
\epsfbox{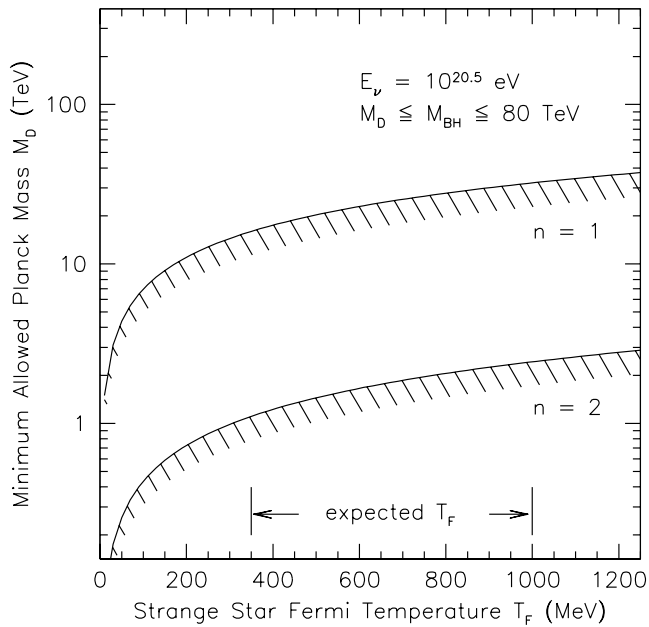}
\caption{Limits on black hole production and the gravity scale
$M_D$ based on bare strange star lifetimes. The limit on $M_D$
is plotted as a function of the assumed Fermi temperature of the
strange star interior $T_F$. Values of 
350 MeV $\leq T_F \leq $ 1000 MeV fall in the acceptable regions
of strange star equations of state, with $T_F \sim 450$~MeV typical. }
\label{tgeom}
\end{center}
\end{figure}

We can place a lower limit on combinations of
$(M_D,~T_F)$ for $n\leq2$ based on the known (at least) several year
lifetimes $\tau_{s}$ of existing strange star candidates. 
We invert equation~(\ref{Enumin}) for 
$E_{\nu}^{max} = 10^{20.5}$~eV, where GZK neutrinos
are expected to have an integral flux comparable to the
measured cosmic ray flux at this energy, 
$F_{\nu}(E_{\nu} \geq 10^{20.5}) \simeq 0.01$~km$^{-2}$~yr$^{-1}$
over $4\pi$ sr. Using this flux, we probe values of 
$M_D$ up to several TeV ($n=2$) or several tens of
TeV ($n=1$). For all values of the neutrino-black hole production
cross section (including exponential suppression) we find that the
implied stellar lifetimes are at most several years, contrary
to observations of the strange star candidates noted previously.

In Fig. 2, we plot the implied limits on 
$(M_D,~T_F)$ based on probable strange star lifetimes greater
than these predictions. There is no real dependence on 
$M_{\rm BH}$ in our analysis, since we find that, as long as the
initial black hole mass is kinematically allowed, it rapidly
acquires all the initial mass-enegy of the incoming neutrino.
Thus our results apply to black hole masses up to $\sim 80$~TeV. Limits
are plotted as a function of $T_F$ the effective thermal bath
temperature of the stellar interior. 

Although the standard spin-down lifetimes that can be estimated for
isolated radio pulsars are not available for strange star candidates, 
their probable lifetimes are at least $10^{6}$~yr, based on
the fact that several of the candidates are in evolved
binary systems. Thus
the lower limit for $M_D$ based on GZK neutrino fluxes is 
constrained only by the lack of fluxes at higher neutrino energies.

Other limits from astrophysical considerations are in fact more stringent
that our limits, but depend on arguments that require analysis of
stellar energy loss due to Kaluza-Klein (KK) gravitons~\cite{Cul99,Bar99}, 
or limits on the decay products of these particles~\cite{Han02}. 
Such KK modes may either be strongly suppressed~\cite{Dval01}, or the
decays may proceed through invisible channels~\cite{Kal00}.
The present results do not depend in any way on the KK emission or
decay processes. The black hole formation process is dependent only 
on geometric and kinematic arguments regarding the extra dimensions,
and further work on understanding the potential stabilization and
growth requirements for these black holes may still extend the 
application or limits to higher values of $n$.

We have focused our discussion on the apparent limits
that obtain for $M_D$ given that strange star candidates appear
to be relatively stable. However, it is interesting to speculate on whether
seeding of such stars with black holes formed via interactions
of ultra-high energy cosmic rays and neutrinos could account to
the observed rate of Gamma-ray burst (GRB) events, since stellar collapse
to a black hole is one mechanism that appears capable of 
producing the observed energy of GRBs events. In fact, in the absence
of other constraints on $M_D$, it is clearly possible to 
create almost any desired rate for stellar collapse of bare
strange stars, since one can almost arbitrarily reduce the cross section
and GZK neutrino fluxes by raising the minimum value for
$M_D$ or $M_{\rm BH}$. The addition of a thin layer of crust material
allows for even further tuning, and thus it appears that black hole
seeding of strange stars as a possible channel for stellar 
collapse is difficult to rule out at present.

We thank Haim Goldberg, Sandip Pakvasa, and Xerxes Tata for
useful discussion and comments on the manuscript. This work was supported
in part by the U.S. Department of Energy Division of High Energy Physics.

\end{document}